\documentclass[preprint,showpacs,amsmath,amssymb,showkeys]{revtex4}

\usepackage{graphicx} 
\usepackage{dcolumn} 
\usepackage{bm} 

\begin{document}

\preprint{}

\title{Strong coupling between mechanical modes in a nanotube resonator}
\author{A. Eichler$^1$}
\author{M. del {\'A}lamo Ruiz$^1$}
\author{J. A. Plaza$^2$}
\author{A. Bachtold$^1$}
\affiliation{$^1$Institut Catal{\`a} de Nanotecnologia, Campus de la UAB, E-08193 Bellaterra, Spain}
\affiliation{$^2$IMB-CNM (CSIC), E-08193 Bellaterra, Barcelona, Spain}
\date{\today}

\begin{abstract}
We report on the nonlinear coupling between the mechanical modes of a nanotube resonator. The coupling is revealed in a pump-probe experiment where a mode driven by a pump force is shown to modify the motion of a second mode measured with a probe force. In a second series of experiments, we actuate the resonator with only one oscillating force. Mechanical resonances feature exotic lineshapes with reproducible dips, peaks, and jumps when the measured mode is commensurate with another mode with a frequency ratio of either $2$ or $3$. Conventional lineshapes are recovered by detuning the frequency ratio using the voltage on a nearby gate electrode. The exotic lineshapes are attributed to strong coupling between the mechanical modes. The possibility to control the strength of the coupling with the gate voltage holds promise for various experiments, such as quantum manipulation, mechanical signal processing, and the study of the quantum-to-classical transition.
\end{abstract}


\maketitle


The nonlinear nature of mode coupling lies at the origin of a wide variety of phenomena~\cite{Sato2003,Cross2004,GilSantos2009,Karabalin2009,Westra2010,Mahboob2011,Venstra2011,Karabalin2011,Mahboob2012,Faust2012,Antoni2012,Lulla2012}, including mechanical synchronization, mechanically-induced transparency, and vibration localization. In the case that two modes operate at well separated frequencies, the effect of the coupling is usually modest and the oscillators move essentially in an independent manner. The coupling between two modes is expected to become strong when the ratio between their resonance frequencies is an integer $n$~\cite{Nayfeh1979}. Perturbation theory then predicts that the motion of one oscillator strongly affects the motion of the other oscillator, and \textit{vice-versa}, through nonlinear forces of order $n$ (see below). Moreover, the resonance lineshapes are expected to be peculiar~\cite{Nayfeh1979}. Strong coupling, also called internal resonance, has not been observed in nanomechanical resonators thus far, because the ratio between the resonance frequencies is usually not an integer.

Resonators based on carbon nanotubes~\cite{Sazonova2004,Lassagne2009,Steele2009} provide a unique platform to test mechanics at the nanoscale. A nanotube behaves like a semi-flexible polymer in the sense that it can bend and stretch to large extents~\cite{Salvetat1999}. Consequently, nonlinearities in nanotube resonators are important and result in unusual behaviours~\cite{Eichler2011,Atalaya2011,Barnard2011}. In this work, we take advantage of the mechanical flexibility of nanotubes to achieve strong coupling between mechanical modes. Indeed, the static shape of a nanotube can be deformed to a large extent with the voltage applied on a gate electrode. This enables us to tune the resonance frequencies~\cite{Sazonova2004,Eichler2011B} in order to make two modes commensurate. In addition, the oscillation is easily driven to large amplitudes~\cite{Sazonova2004} so that nonlinear forces, including coupling forces, are sizeable.

In this Letter, we study the lineshape of mechanical resonances as a function of gate voltage. Lineshapes become exotic, featuring reproducible dips, peaks, and jumps, when the measured mode is commensurate with another mode with a frequency ratio of either $2$ or $3$. Conventional lineshapes are recovered by detuning the frequency ratio with the gate voltage. These results agree with the predictions of strong coupling. The coupling is attributed to motion-induced tension, that is, the oscillation of one mode induces a mechanical tension in the resonator that affects the dynamics of the other mode, and \textit{vice-versa}.


We employ conventional techniques for the fabrication and the measurements of nanotube resonators. Figures~1(a) and (b) show that the nanotube is contacted to two electrodes and is suspended over a trench with a gate electrode at the bottom. The nanotube is grown by chemical vapour deposition in the last step of the fabrication process in order to reduce contamination~\cite{Huttel2009} (supplementary section I). We check with a scanning electron microscope that only one nanotube is suspended over the trench. The mechanical motion is driven and detected using the two-source and the frequency-modulation (FM) mixing methods. The two-source method~\cite{Sazonova2004}, which enables a direct measurement of the amplitude of the motion, is used to record resonance lineshapes, whereas the FM method~\cite{Gouttenoire2010} is better at detecting  small signals so we employ it to map resonance frequencies as a function of gate voltage $V_{g}$ (supplementary section II). Measurements are performed between $60$ and $70$\,K to avoid Coulomb blockade at low temperature~\cite{Lassagne2009,Steele2009}.


The nanotube resonator already begins to exhibit Duffing nonlinearities at low driving force $F_d$~\cite{Postma2005}. Figures~1(c) and (e) show two resonance lineshapes at the lowest $F_d$ for which we obtain a good signal-to-noise ratio. The two resonances correspond to a single mode at different values of $V_{g}$. The quality factors are $230$ and $350$ in Fig.~1(c) and (e), respectively. Upon doubling $F_d$, a hysteresis emerges, marking the onset of the nonlinear regime [Fig.~1(d),(f)]. An estimation of the motional amplitude yields values between $1$\,nm and $9$\,nm in Fig.~1(c-f) (supplementary section III). Interestingly, the asymmetry of the resonance is different between Fig.~1(d) and (f), which indicates different signs of the Duffing force. The sign change occurs around $V_{g} = 1.9$\,V.

The resonance frequencies can be tuned with $V_{g}$ by an amount that is different for each mode [Fig.~2(a),(b)]. The resonance frequency variation is attributed to the mechanical tension that builds up in the nanotube as it bends towards the backgate upon increasing $V_{g}$~\cite{Sazonova2004,Chen2009}. The amount of the variation depends on the shape and the direction of the mode. Finite element simulations can qualitatively reproduce the measured $V_g$ dependences of the different resonance frequencies [Fig.~2(c)] without any free parameters using the static shape of the nanotube imaged with a scanning electron microscope (supplementary section VII). These simulations show that the static deformation of the nanotube towards the gate electrode is as large as $~50$\,nm for $V_{g}=4$\,V [Fig.~2(d)]. For the simpler case of a straight nanotube, we can describe the $V_g$ dependences of the resonance frequencies in a satisfactory way using the Euler-Bernoulli equation (supplementary section VIII); the static deformation is $~17$\,nm for $V_{g}=4$\,V [Fig.~2(e)].

Coupling between the modes can be observed in a pump-probe experiment~\cite{Westra2010}. Specifically, we apply a force at frequency $f_{probe}$ to probe one mode using the FM method. The current of the probed mode is continuously monitored while sweeping the frequency $f_{pump}$ of a second force [Fig.~3(a),(b)]. The sweep in $f_{pump}$ is repeated for various values of $V_{g}$ [Fig.~3(c)]. The current of the probed mode is found to change when $f_{pump}$ matches the resonance frequency (or the harmonic) of another mode [by comparing Fig.~3(c) and supplementary Fig.~S3(a)]. This unambiguously demonstrates that the modes of our nanotube resonator are coupled.

When only one mode is actuated, we observe discontinuities in maps of the resonance frequency as a function of $V_{g}$ [Fig.~4(e),(f)]. The discontinuities are accompanied by exotic resonance lineshapes [Fig.~4(g-j)]. These features often occur when the resonance frequency of the measured mode is equal to that of another mode multiplied by $2$, $3$, $1/2$, or $1/3$ (as indicated by the black arrows in Fig.~2(b);~\cite{remark1}). Upon detuning the frequency ratio using $V_{g}$, conventional resonance lineshapes are recovered [Fig.~4(a),(d)]. Another way to retrieve regular lineshapes is to reduce the driving force (supplementary section XIII). We also observe exotic lineshapes [indicated by gray arrows in Fig.~2(b)] without being able to identify the second mode; we speculate that the second mode is not detectable with the mixing technique or that it oscillates in a frequency range that has not been probed.

These experimental findings are consistent with the theory of strong coupling between mechanical modes in a resonator~\cite{Nayfeh1979,remark2}. The observation that strong coupling occurs for a frequency ratio of $2$ or $3$ implies that quadratic and cubic nonlinear forces are important and that the equation of motion for mode $i$ is of the form

\begin{align}
\frac{d^2 z_i}{d t^2} = - \omega_i^2 z_i - \gamma \frac{d z_i}{d t} - \alpha_2 z_i^2 - \alpha_3 z_i^3 - \sum_{j,k} \beta_{jk} z_j z_k - \sum_{j,k,l} \epsilon_{jkl} z_j z_k z_l + g
\end{align}

with $z_i$ the motional amplitude, $t$ the time, $\omega_i$ the angular resonance frequency, and $g$ the effective force normalized by the mass~\cite{Nayfeh1979}; $\gamma$, $\alpha_2$, $\alpha_3$, $\beta_{jk}$, and $\epsilon_{jkl}$ are various constants. We omit the nonlinear damping force for simplicity~\cite{Eichler2011}. Mode $i$ couples to modes $j$, $k$, and $l$ through the forces $z_j z_k $ and $z_j z_k z_l$ (supplementary section IX).

Quadratic and cubic nonlinear forces ($z_i^2$, $z_i^3$, $z_j z_k $ and $z_j z_k z_l$) naturally emerge from the tension in the beam that is induced by motion - the beam is stretched and compressed periodically in time because it is clamped at both ends. The $z_i^2$ and $z_i^3$ forces are responsible for the hystereses and the asymmetric resonance lineshapes in Fig.~1(d) and (f). The upward asymmetry in Fig.~1(d) is associated with the cubic $z_i^3$ force, since motion-induced tension leads to a positive coefficient $\alpha_3$. When the static deformation of the beam $z_{s}$ becomes sizeable, the quadratic $z_i^2$ force can lead to a reversal of the asymmetry~\cite{Dykman1971,Kozinsky2006}. We estimate from the asymmetries in Fig.~1(d) and (f) that $z_{s}$ is $2.8$ and $13$\,nm at $V_g = 1.5$ and $4$\,V, respectively (supplementary section XI). This is in fair agreement with the calculation in Fig.~2(e), which supports that the nonlinear $z_i^2$ and $z_i^3$ forces originate from motion-induced tension. We estimate that these forces are $3$ orders of magnitude larger than electrostatic nonlinear forces~\cite{Kozinsky2006} and thus neglect the latter (supplementary section XII). The coupling forces $z_j z_k $ and $z_j z_k z_l$ are intimately related to the $z_i^2$ and $z_i^3$ forces, since they all arise in the same way from the Euler-Bernoulli equation (supplementary section IX). It is thus likely that the modal coupling in our experiment is also due to motion-induced tension. In other words, the coupling is mediated by the tension generated by the oscillation of one mode, which affects the dynamics of the other mode, and \textit{vice-versa}. The solutions of the equations of motion that describe motion-induced tension [Eq.~(1)] are characterized by exotic lineshapes for the case of commensurable resonance frequencies~\cite{Nayfeh1979}. The lineshapes are sensitive to the coefficients of the coupling forces in a critical fashion. A detailed comparison between experiment and theory is not possible at the moment, since the coefficients depend on the static shape of the nanotube, which is not known precisely enough.

The exotic lineshapes in nanotube resonators are analogous to Fermi resonances observed in the infrared and Raman spectra of molecules~\cite{Fermi1931,Herzberg1956}. When the frequency of a vibrational mode of a molecule is twice as large as that of another mode, energy can be transferred from one mode to the other. This leads to a mixing of the eigenfunctions and to unusual spectra. However, the coupling between the vibrational modes cannot be externally tuned as in nanotube resonators.

The mode coupling force can be made larger in nanotube resonators than in resonators made from other materials, since the coupling force scales inversely with the fourth power of the resonator length (supplementary section IX) and that nanotube resonators can be as short as $\sim100$\,nm~\cite{Chaste2011,Laird2012}. Mode coupling is further enhanced by the excellent material characteristics of nanotubes, since the coupling force is linearly proportional to $E/\rho$ (supplementary section IX) and that nanotubes have a high Young modulus $E$ and a low mass density $\rho$.


The achievement of strong coupling combined with the possibility to tune its strength open up many possibilities. Such coupling may lead to sizeable signatures in the quantum-to-classical transition of a mechanical resonator~\cite{Katz2007}. In the quantum regime, it may allow for the manipulation of energy quanta between different mechanical modes using gate voltage pulses. Classically, the transfer of energy between mechanical modes could be made faster than the energy relaxation time, which is interesting for high-speed signal operation~\cite{Liu2008,Unterreithmeier2010,Mahboob2012,Yamaguchi2012}. The nonlinear nature of strong coupling is expected to give rise to non-intuitive behaviours that have not been tested thus far~\cite{Nayfeh1979}. A striking example is that driving one of two coupled modes can cause the second mode to reach a higher amplitude than that of the driven one.

When finalizing the manuscript, we became aware of the paper by Antonio et al.~\cite{Antonio2012} that reports on strong coupling in a $0.5$\,mm long micromechanical resonator. The frequency of the modes is tuned by increasing the driving force (through the Duffing force). The possibility to tune the resonance frequencies of a nanotube resonator with a gate voltage is more convenient for practical use.

We thank M. Dykman, A. Isacsson, and H. Yamaguchi for discussions. We acknowledge support from the European Union through the RODIN-FP7 project, the ERC-carbonNEMS project, and a Marie Curie grant (271938), the Spanish ministry (FIS2009-11284), and the Catalan government (AGAUR, SGR). The ANSYS simulation software was financially supported by the MINAHE3 and MINAHE4 projects (Ref. TEC2008-06883-C03-01 and TEC2011-29140-C03-01). We thank Brian Thibeault (Santa Barbara) for help in fabrication.

\newpage

\begin{figure}
\includegraphics[width=86mm]{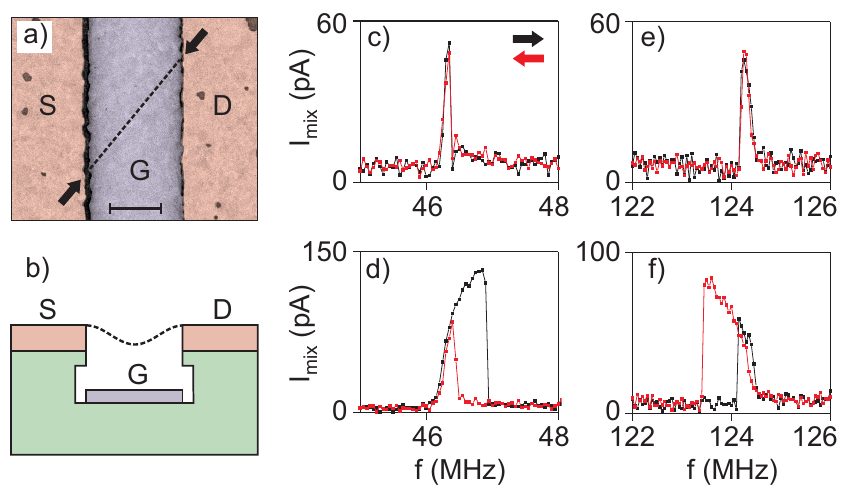}
\caption{\label{Figure 1} Device characterization. (a) Coloured scanning electron microscopy image of the device measured in this work with source (S), drain (D), and gate (G) electrodes. The nanotube position is represented by a dashed line and the clamping points are indicated by arrows. The suspended length of the nanotube is $1.77\,\mu$m and the depth of the trench is $370$\,nm. Scale bar: $600$\,nm. The image is recorded after the measurements of the resonator. (b) Schematic sideview of the device. (c-f) Mechanical resonances for small and large driving forces obtained by measuring the mixing current ($I_{mix}$) as a function of the driving frequency ($f$) with the two-source technique. The driving force is electrostatic and is proportional to the oscillating voltage $V^{ac}$ applied to the gate electrode. $V^{ac} = 0.2$\,mV and $V_{g} = 1.5$\,V in (c); $V^{ac} = 0.4$\,mV and $V_{g} = 1.5$\,V in (d); $V^{ac} = 0.2$\,mV and $V_{g} = 4$\,V in (e); $V^{ac} = 0.4$\,mV and $V_{g} = 4$\,V in (f). For the detection, we apply an oscillating voltage ($V_{s}^{ac} = 0.056$\,mV) to the source electrode. Black (red) curves correspond to upward (downward) sweeps.}
\end{figure}

\clearpage

\begin{figure*}
\includegraphics[width=172mm]{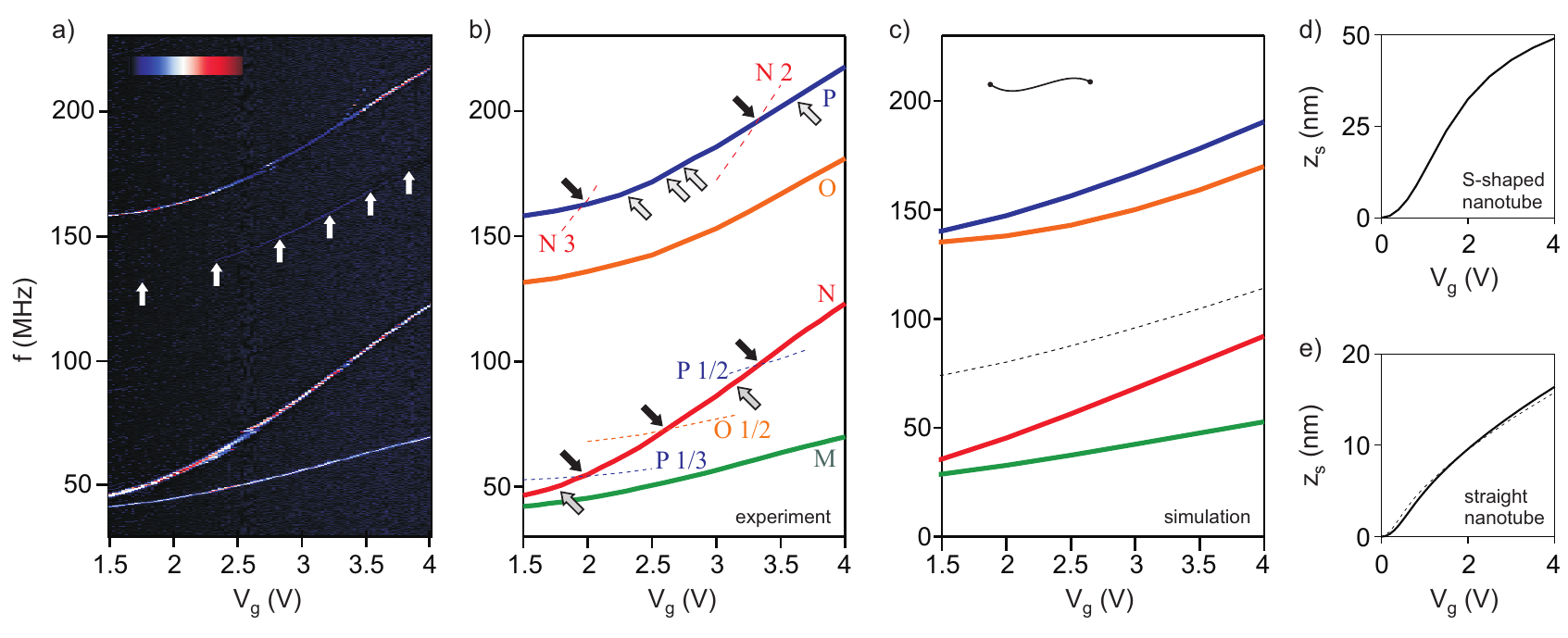}
\caption{\label{Figure 2} Tuning resonance frequencies. (a) Map of resonance frequencies as a function of $V_g$ (by measuring $I_{mix}$ as a function of $f$ and $V_{g}$ for $V^{ac} = 2$\,mV with the FM technique). We clearly discern three modes, while a fourth one is weaker and is indicated by arrows. Colour scale: $0$ (black) to $0.1$\,nA (red). (b) Schematic of the map of resonance frequencies as a function of $V_g$. The four modes are represented by plain lines and labelled $M$, $N$, $O$, and $P$. Dashed lines correspond to the resonance frequencies of these modes multiplied by $2$, $3$, $1/2$, or $1/3$ (the values are indicated in the labels). Black arrows point to regions where lineshapes are exotic and two modes are commensurate. Grey arrows point to exotic lineshape regions for which we cannot assign the coupled mode. (c) Finite element simulation of the map of resonance frequencies as a function of $V_g$ obtained with ANSYS. The dashed line correspond to a mode that we have not detected. The simulations show that this mode has one node (while the others have either zero or two nodes) and thus cannot be detected due to symmetry reasons. Inset: schematic of the static shape of the nanotube when $V_g = 0$\,V. The deformation in the transverse direction is exaggerated with respect to the nanotube length. The largest deformation is $\sim 40$\,nm (supplementary section VII). (d) Static displacement of the center of the resonator ($z_s$) calculated with ANSYS using the static shape of the nanotube (when $V_g = 0$\,V) depicted in the inset of (c). (e) $z_s$ calculated from the Euler-Bernoulli equation for a straight nanotube (plain line). The dashed line corresponds to the result calculated with ANSYS for the same straight nanotube.}
\end{figure*}

\clearpage

\begin{figure}
\includegraphics[width=86mm]{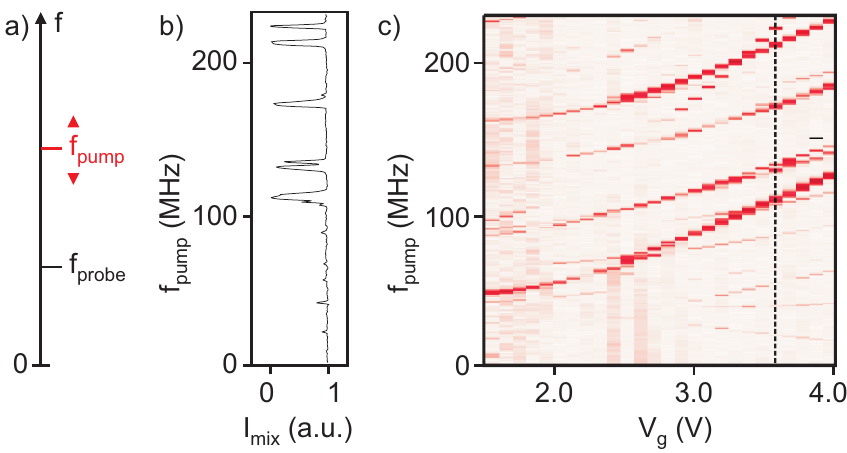}
\caption{\label{Figure 3} Mechanical coupling measured in a pump-probe experiment. (a) Representation of the two drive frequencies used in the pump-probe experiment. The pump force is swept in frequency, whereas $f_{probe}$ is set to match the resonance frequency of the lowest mode ($M$). (b) Normalized mixing current of the probed mode as a function of $f_{pump}$ at $V_{g} = 3.6$\,V (measured with the FM technique). Before the scan, we set $f_{probe}$ so that the current is maximal ($I_{mix}^0$). We plot the measured current divided by $I_{mix}^0$. The oscillating voltage of the pump is $5.6$\,mV and the FM oscillating voltage of the probe is $2$\,mV. (c) Normalized mixing current of the probed mode as a function of  $f_{pump}$ and $V_g$ using the same parameters as in (b). The line graph in (b) is marked with a dashed line. Colour scale: $I_{mix}=0$ (dark red) to $I_{mix}=1$ (white).}
\end{figure}

\clearpage

\begin{figure}
\includegraphics[width=86mm]{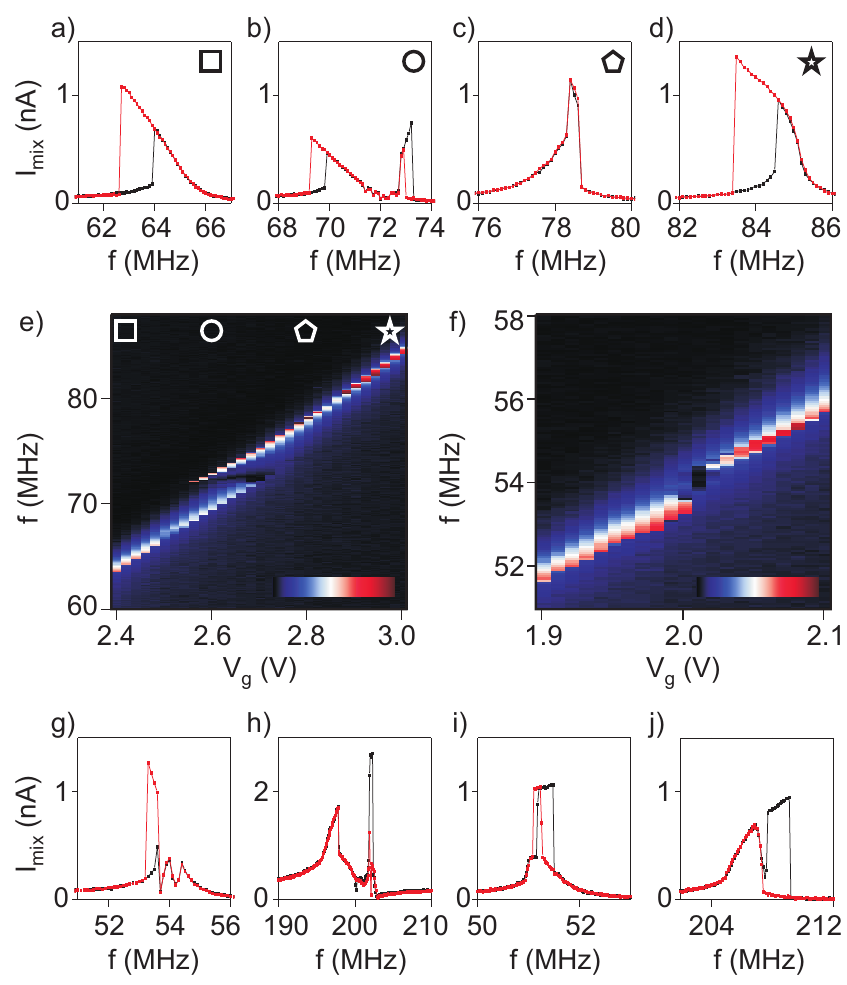}
\caption{\label{Figure 4} Resonances when the measured mode is commensurate or nearly commensurate with another mode. (a-d) Resonance lineshapes of one mode for different $V_g$ measured with the two-source technique. The driving voltage applied to the gate electrode is $V^{ac} = 1.1$\,mV and the voltage applied for the detection on the source electrode is $V_{s}^{ac} = 0.56$\,mV. Black (red) curves correspond to upward (downward) sweeps. $V_g = 2.4$\,V in (a); $V_g = 2.6$\,V in (b); $V_g = 2.8$\,V in (c); $V_g = 3$\,V in (d); (e) Map of the resonance frequency as a function of $V_g$ (obtained by measuring $I_{mix}$ as a function of $f$ and $V_g$ for $V^{ac} = 1.1$\,mV and $V_{s}^{ac} = 0.56$\,mV using the two-source technique by increasing $f$). Colour scale: $0$ (black) to $1$\,nA (red). (f) Map of the resonance frequency as a function of $V_g$ (obtained for $V^{ac} = 1.1$\,mV and $V_{s}^{ac} = 0.28$\,mV using the two-source technique by decreasing $f$). Colour scale: $0$ (black) to $0.7$\,nA (red). (g-j) Resonance lineshapes for different modes and different values of $V_g$ measured with the two-source technique. $V^{ac} = 1.1$\,mV, $V_{s}^{ac} = 0.56$\,mV, and $V_g = 1.98$\,V in (g). $V^{ac} = 17$\,mV, $V_{s}^{ac} = 1.1$\,mV, and $V_g = 3.41$\,V in (h).  $V^{ac} = 1.7$\,mV, $V_{s}^{ac} = 0.28$\,mV, and $V_g = 1.86$\,V in (i). $V^{ac} = 5.6$\,mV, $V_{s}^{ac} = 0.56$\,mV, and $V_g = 3.6$\,V in (g).}
\end{figure}

\clearpage

\large
\begin{center}

\textbf{Supplementary Material for: Strong coupling between mechanical modes in a nanotube resonator}

\normalsize

\vspace{5 mm}

A. Eichler$^1$, M. del {\'A}lamo Ruiz$^1$, J. A. Plaza$^2$, and A. Bachtold$^1$

\textit{$^1$Institut Catal{\`a} de Nanotecnologia, Campus de la UAB, E-08193 Bellaterra, Spain and}

\textit{$^2$IMB-CNM (CSIC), E-08193 Bellaterra, Barcelona, Spain}

\end{center}

\small

\section{Device fabrication}
Our nanoresonators consist of a suspended carbon nanotube clamped between two metal electrodes, as depicted in Fig.~1(a) and (b) of the main text. The devices are fabricated as follows. A trench is etched into a highly resistive Si wafer coated with SiO$_2$ and Si$_3$N$_4$. W and Pt are evaporated into the trench to create a gate electrode (G). In a second lithography step, a continuous line is exposed across the trench. After a deposition of W/Pt and lift-off, the line results in the source (S) and drain (D) electrodes separated by the trench (these electrodes are electrically isolated from the gate due to the undercut profile of the Si$_3$N$_4$/SiO$_2$ substrate). W and Pt are chosen because of their high melting points that allow the growth of carbon nanotubes. An island of catalyst is patterned on the drain (or source) electrode using electron-beam lithography. Nanotubes are grown by chemical vapour deposition from these islands. In about $1$ out of $20$ cases, a nanotube grows across the trench and establishes electrical contact between S and D. This growth is the last step of the fabrication process so that nanotubes are not contaminated with residues from resists and chemicals~\cite{Cao2005S,Huttel2009S,Steele2009S}. The device we present in the main text has a length of $1.77$\,$\mu$m. The separation between the nanotube and the gate electrode is $370$\,nm. The rather large roughness of the S and D electrodes in this device does not allow us to measure the nanotube radius with atomic force microscopy.

\section{Measurements details}
Our measurements are carried out at pressures typically below $10^{-8}$\,mbar and temperatures between $60$ and $70$\,K. In order to clean the nanotube surface, we perform a current annealing step every day ($6$\,$\mu$A for $300$\,s). We observe only very minor variations of the electrical conductance and mechanical resonance frequencies of the nanotube from day to day.

The chip containing the device is mounted on a printed circuit board. dc and ac voltages are added through a bias tee outside the chamber. The low frequency mixing current is measured from the drain (D) electrode and is low-pass filtered through a capacitor to ground ($1$\,nF).

We discuss first the frequency mixing (FM) technique~\cite{Gouttenoire2010S}. A driving voltage $V^{ac}$ is applied to the source electrode. Modulating the frequency (with a modulation rate of $671$\,Hz and a frequency deviation of $100$\,kHz) results in a mixing current ($I_{mix}$) at $671$\,Hz. The gate electrode is biased with a dc voltage $V_g$ to tune the resonance frequencies.

In the two-source technique~\cite{Sazonova2004S}, we apply the driving voltage $V^{ac}$ to the gate in addition to a dc voltage $V_g$. The motion of the nanotube is detected by applying a second, smaller voltage $V_s^{ac}$ to the source. The two oscillating voltages are slightly detuned, and the amplitude signal of $I_{mix}$ is measured at the detuning frequency ($\delta \omega / 2 \pi = 10$\,kHz).

The modulus of $I_{mix}$ measured with the two-source technique has the form

\begin{equation}
I_{mix} = \frac{1}{2} V_{s}^{ac} \frac{\partial G}{\partial V_g}\left(V^{ac} \cos(\delta \omega t - \varphi_E) + z_0 V_g \frac{C'}{C} \cos(\delta \omega t - \varphi_E - \varphi_M) \right) \label{mixingcurrent}
\end{equation}

where $G$ is the conductance of the nanotube, $\varphi_E$ is the phase difference between the voltages applied to source and gate, $t$ is time, $z_0$ is the mechanical amplitude, $C$ is the capacitance between the nanotube and the gate, $C'$ is its derivation with respect to the nanotube displacement, and $\varphi_M$ is the  phase difference between the nanotube displacement and the driving force.

The measurements in the paper of $I_{mix}$ as a function of the drive frequency $f$ give a resonance lineshape that is to a rather good approximation  proportional to the response of the motional amplitude as a function of $f$, since the purely electrical component of $I_{mix}$ (first term in Eq.~\ref{mixingcurrent}) is much lower than the mechanical component (second term in Eq.~\ref{mixingcurrent}).

We verify that the harmonics of the RF sources (signals at $2$, $3$, or $1/2$ times the drive frequency) can be neglected. These harmonics are far below the smallest driving voltage for which we can detect a resonance. Namely, the voltage of the harmonics is typically $1000$ times lower than $V^{ac}$.

\section{Estimation of dynamical amplitude}
Equation~\ref{mixingcurrent} allows estimating the motional amplitude of the resonator by comparing the signal on resonance, $I_{max}$, to the purely electrical background far from resonance, $I_{back}$~\cite{Sazonova2004S}. Using the approximation $C = \frac{2 \pi \epsilon_0 L}{\ln(2 (d - z) / r)}$ , we get that

\begin{equation}
z_0 \simeq d \cdot \ln\left(\frac{2 d}{r} \right) \frac{I_{max}}{I_{back}} \frac{V^{ac}}{V_g}
\end{equation}

with $d = 370$\,nm the equilibrium distance between the nanotube and the gate electrode. Since we cannot measure the diameter of the nanotube due to the large surface roughness of the electrodes in the studied device, we use a typical value for the radius ($r = 1.5$\,nm). We find the following values for the maximum mechanical amplitudes $z_0$ in Fig.~1 of the main text: $z_0 \simeq 3.2$\,nm in Fig.~1(c), $z_0 \simeq 8.7$\,nm in Fig.~1(d), $z_0 \simeq 0.9$\,nm in Fig.~1(e), and $z_0 \simeq 2.1$\,nm in Fig.~1(f).

\begin{figure*}
\includegraphics[width=172mm]{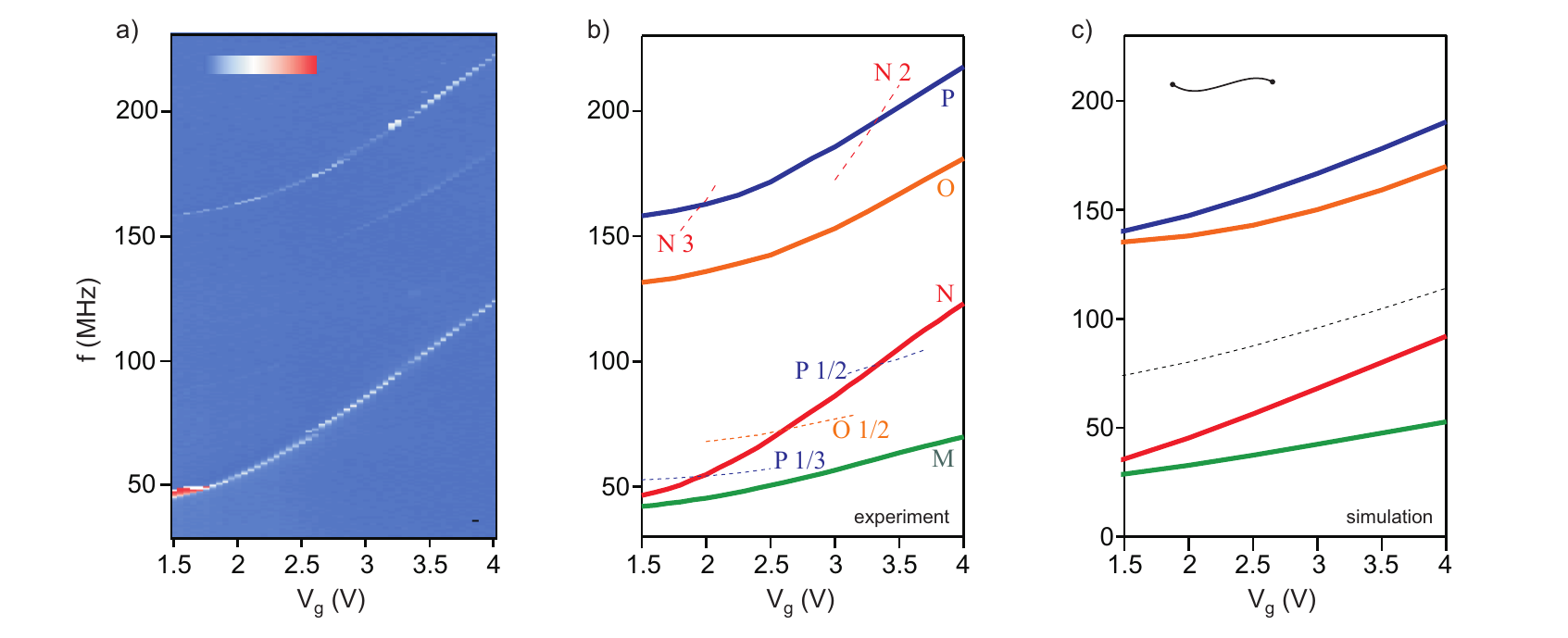}
\caption{\label{SM13} Maps of resonance frequencies as a function of gate voltage. (a) Two-source measurement at low driving force (obtained by measuring $I_{mix}$ as a function of $f$ and $V_g$ with $V^{ac}=1.7$\,mV and $V_{s}^{ac} = 0.3$\,mV). Colour scale: $0$ (blue) to $1$\,nA (red). For comparison, we plot the schematic of the modes detected with the FM technique in (b) and the results of the ANSYS simulation in (c). The inset in (c) shows the static nanotube shape (measured in the absence of an applied dc voltage) that is used in the simulation.}
\end{figure*}

\section{Maps of resonance frequencies as a function of gate voltage: comparison between the two-source and the FM techniques at low driving forces}
We carry out measurements of the resonance frequencies as a function of $V_g$ with the two-source technique at low driving force [Fig.~\ref{SM13}(a)]. Two modes are detected with a large signal, a third one with a small signal. To facilitate a comparison between the two measurement techniques, we plot the schematic of the modes detected with the FM technique in Fig.~\ref{SM13}(b). We find that the modes producing a large signal with the two-source technique are those we label $N$ and $P$. The mode producing a weak signal is identified as $O$, while $M$ is not detected at all (it does show up at larger driving forces).

The ANSYS simulation helps understanding the relative strengths of the signals [Fig.~\ref{SM13}(c)]. The details of the simulation are discussed in section VII. $N$ and $P$ correspond to modes moving essentially in the plane orthogonal to the gate electrode. Since the two-source method measures the oscillation of the nanotube-gate capacitance, the signal of modes $N$ and $P$ are expected to be large, in agreement with the experiments. $M$ and $O$ are modes moving essentially parallel to the gate electrode. As such, the corresponding signals are expected to be small, which also agrees with the measurement.

ANSYS simulations indicate that modes $M$, $N$, $O$, and $P$ have either $0$ or $2$ nodes. The mode with $1$ node moving perpendicular to the gate electrode is predicted to appear between $N$ and $O$ [dashed line in Fig.~\ref{SM13}(c)]. This mode is not detected because the oscillation of the capacitance is (nearly) zero due to the symmetry of the mode shape. ANSYS predicts that the mode with $1$ node parallel to the gate has a frequency larger than that of modes $O$ and $P$.

\begin{figure*}
\includegraphics[width=172mm]{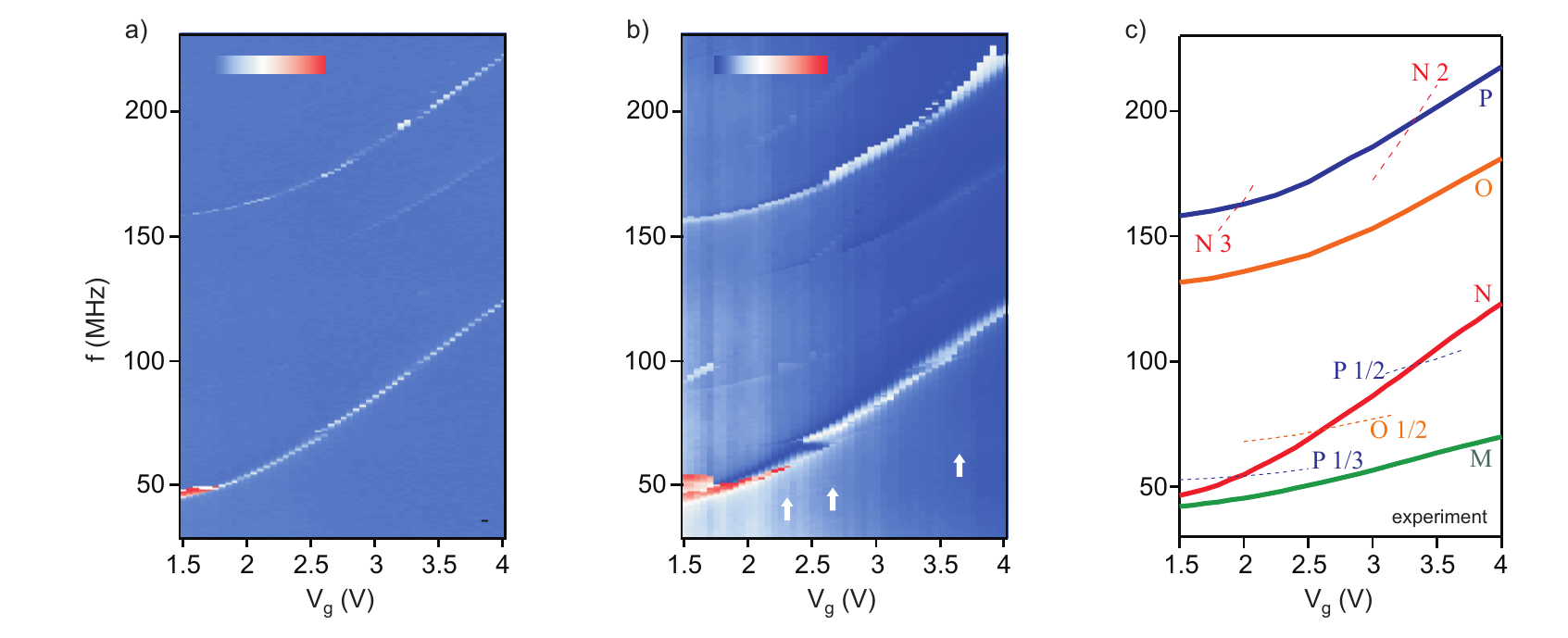}
\caption{\label{SM14} Map of resonance frequencies as a function of gate voltage. (a) Two-source measurement at low driving force (obtained by measuring $I_{mix}$ as a function of $f$ and $V_g$ with $V^{ac}=1.7$\,mV and $V_{s}^{ac} = 0.3$\,mV). Colour scale: $0$ (blue) to $1$\,nA (red). (b) Same measurement with a larger driving force ($V^{ac}=17$\,mV and $V_{s}^{ac} = 1.1$\,mV). White arrows point out the lowest mode which is faintly visible. Colour scale: $0$ (blue) to $7$\,nA (red). (c) Schematic of the modes detected with the FM technique for comparison.}
\end{figure*}

\begin{figure*}
\includegraphics[width=172mm]{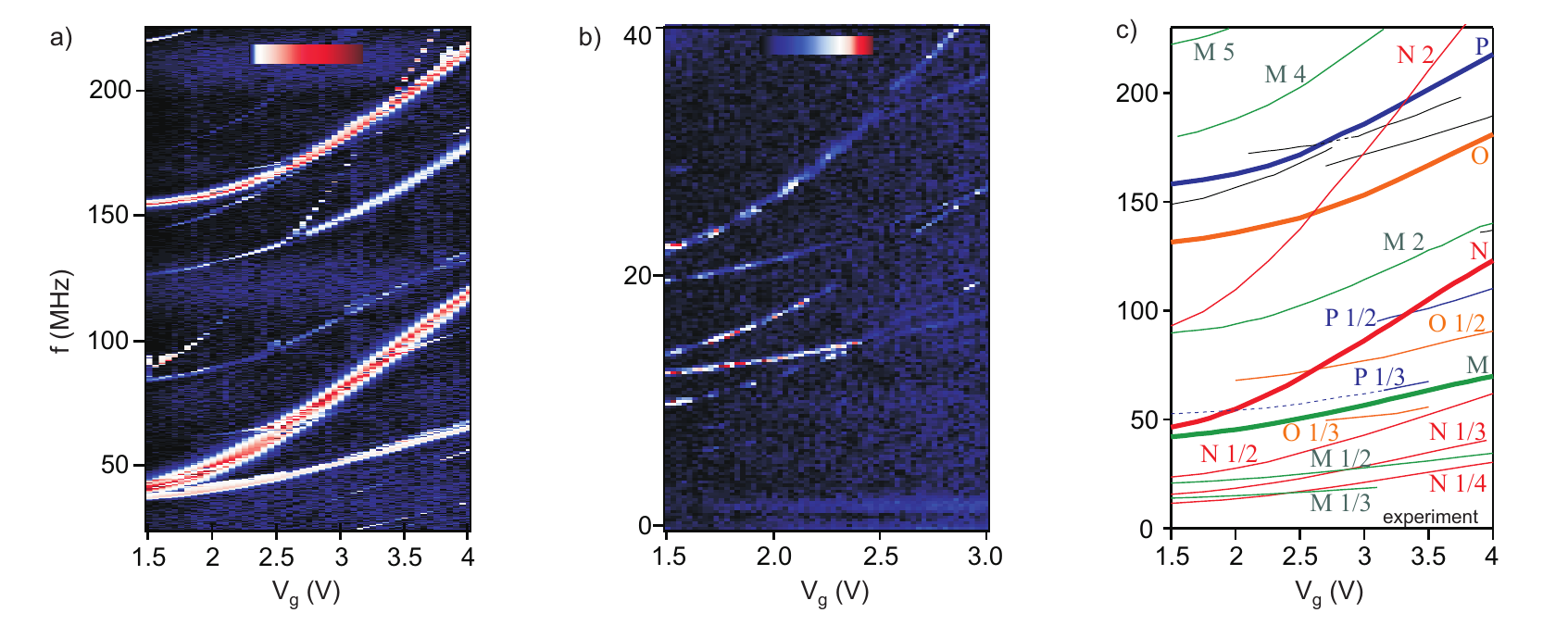}
\caption{\label{SM1} Map of resonance frequencies as a function of gate voltage measured with the FM technique. (a) Frequency modulation measurement at large driving force (obtained by measuring $I_{mix}$ as a function of $f$ and $V_g$ with $V^{ac}=20$\,mV). Colour scale: $0$ (black) to $1$\,nA (dark red). (b) Low frequency range at even higher driving force ($V^{ac}=40$\,mV). Colour scale: $0$ (black) to $0.2$\,nA (dark red). (c) Map of all detected modes and harmonics. The number in the label designates the harmonic order of a resonance. Black lines correspond to resonances that cannot be assigned to a detected mode.}
\end{figure*}

\section{Maps of resonance frequencies as a function of gate voltage: comparison between low and high driving forces}
In the maps of resonances as a function of $V_{g}$ the number of detected resonances depends on the driving force $F_d$. Figure~\ref{SM14}(a) shows the spectrum obtained with the two-source method for a low driving force. Upon increasing $F_d$ by a factor 10, many more resonances appear Fig.~\ref{SM14}`(b)]. A comparison of the resonance frequencies reveals that the additional resonances are almost all harmonics of the four modes $M$, $N$, $O$, and $P$ (see following section).

The same scenario develops for measurements with the FM technique at high driving force: additional resonances are detected [Fig.~\ref{SM1}(a) and (b)] and identified as harmonics of the four modes (see following section). Here, the number of discernible harmonics is even larger than that measured with the two-source technique.

\begin{figure*}
\includegraphics[width=172mm]{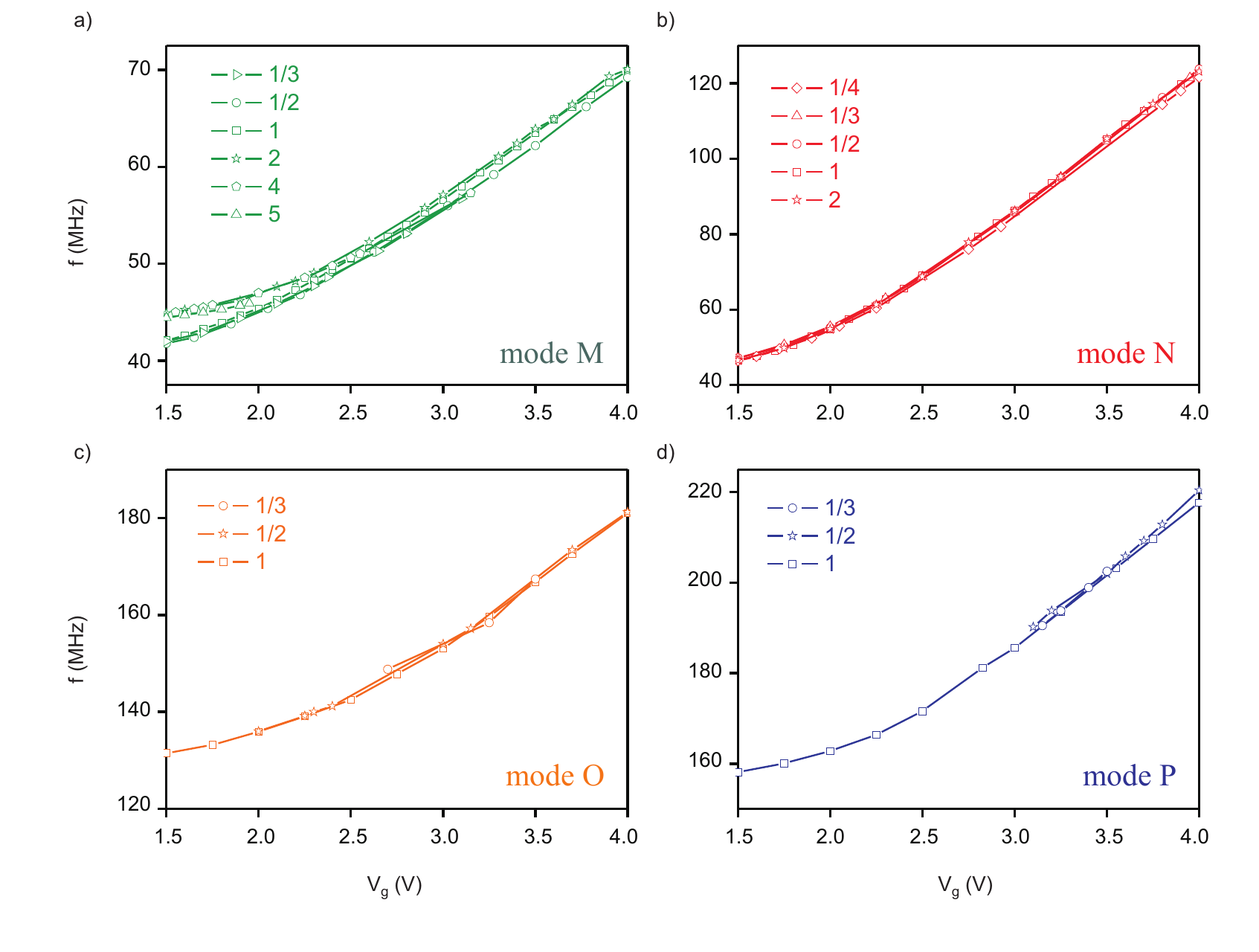}
\caption{\label{SM2} Comparison of the modes with their harmonics. Each harmonic is divided by its respective order [corresponding to the index in Fig.~\ref{SM1}(c)]. (a) Mode $M$ and harmonics. The subharmonics ($M 2$, $M 4$, and $M 5$) deviate slightly at low values of $V_g$. (b) Mode $N$ and harmonics. (c) Mode $O$ and harmonics. (d) Mode $P$ and harmonics.}
\end{figure*}

\section{Harmonics}
In Fig.~\ref{SM2}, we plot the four modes and their harmonics on top of each other by dividing each of them by their respective harmonic order [i.e. the index number in Fig.~\ref{SM1}(c)]. In the case of the modes $N$, $O$, and $P$, the curves are perfectly on top of each other. For mode $M$, the scaling is slightly less good. Harmonics can be generated by several mechanisms. In the following, we will briefly discuss the parametric effect, electrical nonlinearities, and mechanical nonlinearities as possible origin of the harmonics.

A common way to explain harmonics is based on the parametric effect~\cite{Lifshitz2008S,Nayfeh1979S}. A mode at a resonance frequency $f_0$ can be actuated by varying the resonator spring constant $k$ at a frequency $2f_0 / j$, where $j$ is an integer $\geq 1$. It is easy to parametrically drive a nanotube resonator with a gate voltage, because $f_0$ (and therefore $k$) is widely tunable with $V_g$~\cite{Eichler2011BS}. There exists a threshold $V_{th}^{ac}$ above which the motion sets in: this threshold takes the form $V_{th}^{ac} = (f_0 \partial V_g/\partial f_0)/Q$, where $\partial f_0/ \partial V_g$ is the change of the resonance frequency with gate voltage. For mode $N$, we get $\partial f_0/ \partial V_g = 36$\,MHz/V close to $V_{g} = 4$\,V. Together with the quality factor $Q \sim 350$ and $f_0\sim 124$\,MHz, this yields a threshold of $10$\,mV. This is consistent with the harmonics for mode $N$ in Fig.~\ref{SM1}(a) ($V^{ac} = 20$\,mV). However, parametric excitation cannot account for the harmonics of order $4$ and $5$ of mode $M$.

A second scenario for harmonics is related to electrical nonlinearities in the circuit. Nonlinearities in current-voltage characteristics can generate forces at $2$, $3$, $4$,... times the frequency of the applied $V^{ac}$ and thus lead to harmonics with an index number $n < 1$. However, harmonics with an index number $n > 1$ are unlikely to have an electrical origin.

Mechanical nonlinearities are predicted to give rise to harmonics. For instance, the quadratic nonlinear force can cause harmonics at $2 f_0$ and $f_0 / 2$, and the cubic nonlinear force can induce harmonics at $3 f_0$ and $f_0 / 3$~\cite{Nayfeh1979S,Dykman1996S}. The combination of the quadratic and the cubic nonlinear forces can lead to harmonics with index $1/2$, $1/3$, $1/4$, $2$, $3$, $2/3$, $3/2$,...~\cite{Nayfeh1979S}, which is in agreement with our measurements.

In conclusion, the origin of the harmonics is not clear at the moment and this calls for future work.

\section{Simulations}
\begin{figure*}
\includegraphics[width=172mm]{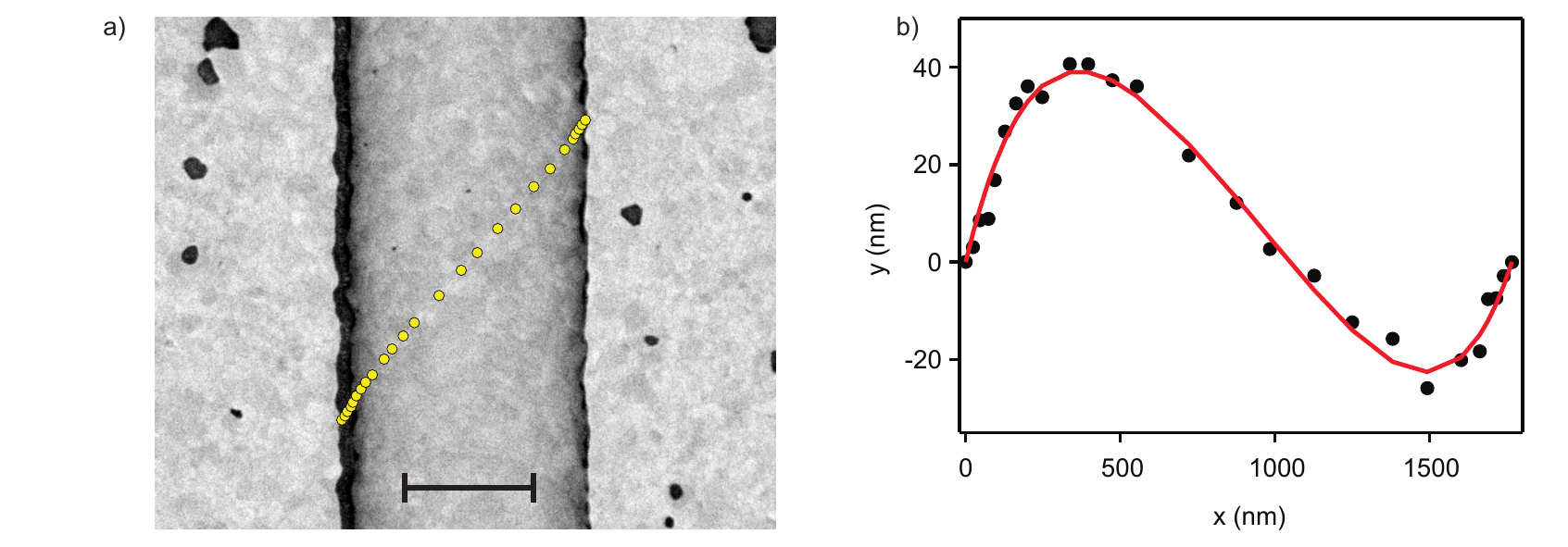}
\caption{\label{SM10} Static shape of the nanotube. (a) Scanning electron micrograph of the nanotube. The electrodes are not voltage biased. Yellow dots mark the data points that we used to model the nanotube shape (for $V_g = 0$\,V). Scale bar: $600$\,nm. (b) Points extracted from the electron micrograph (black dots) and polynomial fit used for the nanotube model (red line). $x$ is the coordinate along the axis connecting the two clamping points. $y$ is the coordinate standing orthogonal to it.}
\end{figure*}

We perform finite element simulations with ANSYS (R) Release 13.0 to reproduce the $V_g$ dependence of the resonance frequencies. The mechanical properties of carbon nanotubes are well described by continuum elasticity and are independent of the chirality~\cite{Kudin2001S}. For these simulations, we use a tube with length $L = 1.77$\,$\mu$m, radius $r = 1.5$\,nm, wall thickness $\Delta r = 0.335$\,nm, mass density $\rho = 2300$\,kg/m$^3$, and Young modulus $E = 1$\,TPa. We use the shape of the nanotube extracted from the scanning electron micrograph of the device (Fig.~\ref{SM10}). We assume that the static deformation is only in the horizontal plane when the device is not voltage biased.

We use the 1-D BEAM188 element suitable for analyzing slender beam structures. 1-D BEAM188 is a two-node element in 3-D and has six degrees of freedom at each node: translations in the $x$, $y$, and $z$ directions and rotations about the $x$, $y$, and $z$ directions. A circular tube section is associated to the element by providing the inner and the outer radii. The constraints at the clamping points are fixed by setting all degrees of freedom to zero. Because of the high aspect ratio of nanotubes, we use point-like clamping conditions. The effect of the angle of the nanotube with respect to the electrodes is thus not accounted for. The electrostatic force induced by the dc voltage applied on the gate electrode, $F_d$, is analytically calculated. For each gate voltage, the static deformation of the nanotube is calculated by performing a nonlinear structural analysis. This static solution is used as a base for the modal analysis.

\section{Euler-Bernoulli equation: mode frequencies}
In the following, we demonstrate that a good qualitative understanding of the motion of a suspended nanotube is possible from the Euler-Bernoulli equation. This approach provides analytical solutions that capture the behaviour of the system and that are similar to the solutions of the finite-element simulations performed with ANSYS.

The Euler-Bernoulli equation for the static and dynamic displacement of a thin beam reads

\begin{equation}
\rho S \frac{d^2z}{dt^2} = -EI \frac{d^4z}{dx^4} + \left[T_0 + \frac{ES}{2L}\int^{L}_{0}\left(\frac{dz}{dx}\right)^2dx\right]\frac{d^2z}{dx^2} + g(t) \label{EulerBernoulli_1}
\end{equation}

where $\rho$ is the mass density, $S$ the beam's cross-sectional area, $z$ the displacement, $t$ the time, $E$ the Young modulus, $I$ the second moment of inertia about the longitudinal axis, $x$ the coordinate along the axis, $T_0$ the built-in tension, $L$ the resonator length, and $g(t)$ a unit length force that accounts for the effect of the gate electrode in our experiment. We divide the displacement into a static and a dynamic component,

\begin{equation}
z(x,t) = z_s \phi_s (x) + z_1(t)\phi_1 (x) \label{EulerBernoulli_2}
\end{equation}

where $z_s$ is the maximum static displacement, $z_1$ is the maximum dynamic displacement, and $\phi_s (x)$, $\phi_1 (x)$ are the normalized static and dynamic profiles along the beam.

In a first example, we develop Eq.~\ref{EulerBernoulli_1} for the case of a single mechanical mode with

\begin{equation}
\phi_s (x) = \phi_1 (x) = \sin(\pi x/L) \label{EulerBernoulli_3}
\end{equation}

where both the static and dynamic profiles are in the plane perpendicular to the gate electrode. This mode profile is strictly correct for negligible bending rigidity ($E I \rightarrow 0$). We choose this ansatz because it allows a simple analytical treatment of our problem. Moreover, we will see at the end of this section that this ansatz predicts gate voltage dependencies of the resonance frequencies that are in qualitative agreement with finite element simulations. We insert Eq.~\ref{EulerBernoulli_2} and Eq.~\ref{EulerBernoulli_3} into Eq.~\ref{EulerBernoulli_1}, multiply Eq.~\ref{EulerBernoulli_1} by $\phi_1 (x)$, and integrate it from $0$ to $L$ to get

\begin{align}
\frac{d^2 z_1(t)}{dt^2} = - \frac{1}{\rho S}\left[E I z_s\left(\frac{\pi}{L} \right)^4 + T_0 z_s \left(\frac{\pi}{L} \right)^2 + \frac{E S}{4} z_s^3 \left(\frac{\pi}{L} \right)^4 - \frac{4}{\pi} g(t) \right] \nonumber \\
- \frac{1}{\rho S} \left[E I \left(\frac{\pi}{L} \right)^4 + T_0 \left(\frac{\pi}{L} \right)^2 + \frac{3}{4} E S z_s^2 \left(\frac{\pi}{L} \right)^4 \right] z_1(t) \nonumber \\
- \left[\frac{3 E}{4 \rho} z_s \left(\frac{\pi}{L} \right)^4 \right] z_1^2(t) - \left[\frac{E}{4 \rho} \left(\frac{\pi}{L} \right)^4 \right] z_1^3(t). \label{EulerBernoulli_4}
\end{align}

In a static equilibrium position, the sum of the static terms in the first bracket on the right hand side of Eq.~\ref{EulerBernoulli_4} is zero:

\begin{equation}
E I z_s\left(\frac{\pi}{L} \right)^4 + T_0 z_s \left(\frac{\pi}{L} \right)^2 + \frac{E S}{4} z_s^3 \left(\frac{\pi}{L} \right)^4 = \frac{4}{\pi} g(t). \label{EulerBernoulli_5}
\end{equation}

The other terms of Eq.~\ref{EulerBernoulli_4} can be rewritten in the usual form of a Newton equation of motion,

\begin{equation}
\frac{d^2z_1(t)}{dt^2} = - \omega_0^2 z_1(t) - \alpha_2 z_1^2(t) - \alpha_3 z_1^3(t). \label{EulerBernoulli_6}
\end{equation}

From a comparison of Eq.~\ref{EulerBernoulli_4} to Eq.~\ref{EulerBernoulli_6}, we see that both $\alpha_2$ and $\alpha_3$ are positive, and that $\alpha_2 \propto z_s$ (meaning that $\alpha_2$ will vanish if the tube is straight). Further, both nonlinear coefficients are inversely proportional to $L^4$ and will become large for a short tube. They arise from additional tension that is generated when the beam bends (the integral term $\int_0^L \left(\frac{d z}{d x} \right)^2 dx$ in Eq.~\ref{EulerBernoulli_1} becomes nonzero).

We repeat this calculation for other modes. The profile of the second mode, $\phi_2 (x)$, has the same shape as $\phi_1 (x)$, but stands orthogonal to it, moving parallel to the gate electrode. In this case, the final equation describing the second mode is somewhat simpler than that for the first mode:

\begin{align}
\frac{d^2 z_2(t)}{dt^2} = - \frac{1}{\rho S} \left[E I \left(\frac{\pi}{L} \right)^4 + T_0 \left(\frac{\pi}{L} \right)^2 + \frac{E S}{4} z_s^2 \left(\frac{\pi}{L} \right)^4 \right] z_2(t) - \left[\frac{E}{4 \rho} \left(\frac{\pi}{L} \right)^4 \right] z_2^3(t). \label{EulerBernoulli_7}
\end{align}

Again, $\alpha_3$ is positive, but here $\alpha_2 = 0$ because the beam features no static bending in the direction of its vibrations. The linear restoring force of $z_1(t)$ is always larger than that of $z_2(t)$, causing $\omega_1 \geq \omega_2$.

We assume that the third and fourth modes have the profile

\begin{equation}
\phi_3 (x) = \phi_4 (x) = \sin(2 \pi x/L), \label{EulerBernoulli_8}
\end{equation}

moving towards and parallel to the gate electrode, respectively. The resonance frequencies of the two modes are degenerate. Neither of these modes are detected in our experiment due to the antisymmetrical mode profile. The solution in this case reads (with $i = 3$ or $4$)

\begin{align}
\frac{d^2 z_i(t)}{dt^2} = - \frac{1}{\rho S} \left[E I \left(\frac{2 \pi}{L} \right)^4 + T_0 \left(\frac{2 \pi}{L} \right)^2 + \frac{E S}{16} z_s^2 \left(\frac{2 \pi}{L} \right)^4 \right] z_i(t) - \left[\frac{E}{4 \rho} \left(\frac{2 \pi}{L} \right)^4 \right] z_i^3(t). \label{EulerBernoulli_9}
\end{align}

We assume that the fifth and sixth modes are analogous to the third and fourth, but with

\begin{equation}
\phi_5 (x) = \phi_6 (x) = \sin(3 \pi x/L). \label{EulerBernoulli_10}
\end{equation}

Again, they are degenerate. We get (with $j = 5$ or $6$)

\begin{align}
\frac{d^2 z_j(t)}{dt^2} = - \frac{1}{\rho S} \left[E I \left(\frac{3 \pi}{L} \right)^4 + T_0 \left(\frac{3 \pi}{L} \right)^2 + \frac{E S}{36} z_s^2 \left(\frac{3 \pi}{L} \right)^4 \right] z_j(t) - \left[\frac{E}{4 \rho} \left(\frac{3 \pi}{L} \right)^4 \right] z_j^3(t). \label{EulerBernoulli_11}
\end{align}

\begin{figure*}
\includegraphics[width=172mm]{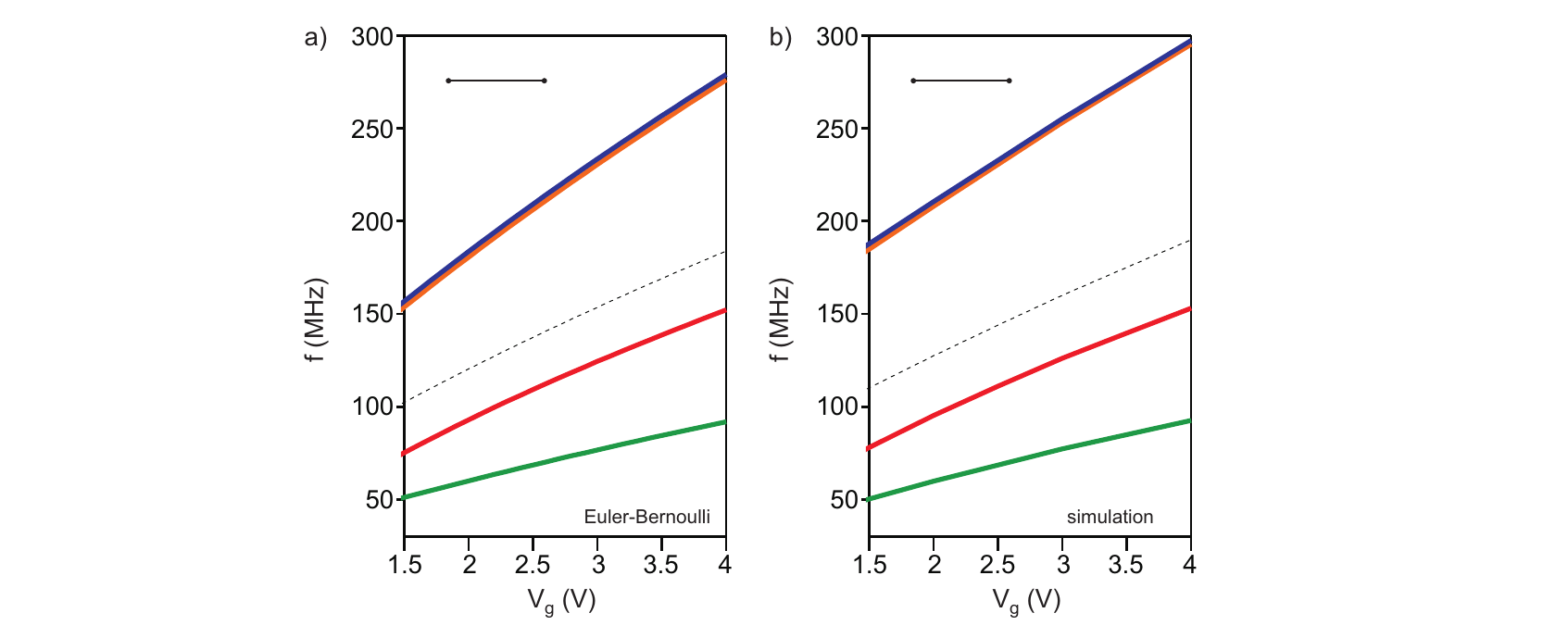}
\caption{\label{SM12} Calculation of resonance frequencies as a function of $V_g$ for a straight nanotube. (a) Resonance frequencies of the first six modes calculated from the Euler-Bernoulli equation. red: $z_1$, green: $z_2$, black dashed: $z_3$ and $z_4$ (not detected in our experiment due to the antisymmetrical mode profile), blue: $z_5$, orange: $z_6$. Inset: shape of the nanotube when $V_g = 0$\,V. (b) ANSYS simulation for the same set of parameters. The results are very similar to the analytical calculations.}
\end{figure*}

At this point, we can calculate the resonance frequencies of the first six modes as a function of $V_g$. For this, we determine the static displacement that provides an equilibrium of forces by solving Eq.~\ref{EulerBernoulli_5}, where 

\begin{align}
g(t) =  \frac{1}{2} c' V_g^2 \label{EulerBernoulli_12}
\end{align}

is the unit length force due to $V_g$, and

\begin{align}
c' = \frac{2 \pi \varepsilon_0}{d \ln(2 d / r)^2} \label{EulerBernoulli_13}
\end{align}

is the differentiation of the unit length capacitance with respect to the displacement. Here, $d$ is the distance between the nanotube and the gate electrode, $\varepsilon_0 = 8.85 \cdot 10^{-12}$\,Fm$^{-1}$ is the electrical permittivity of free space, and $r$ is the nanotube radius. We measure $d = 370$\,nm by atomic force microscopy (AFM), but the large surface roughness of the electrodes of this device does not allow the measurement of $r$. We therefore chose a typical value from earlier nanotubes grown by the same method ($r = 1.5$\,nm). We also use $E = 1$\,TPa, $\rho = 2300$\,kg/m~\cite{Lee2008S}, $L = 1.77$\,$\mu$m (measured by AFM), and wall thickness $\Delta r = 0.335$\,nm. From these values, we calculate the second moment of inertia $I = 2.928 \cdot 10^{-38}$\,kg\,m$^{2}$ and the tubular cross-section $S = \pi ((r + \Delta r)^2 - (r - \Delta r)^2) = 3.157 \cdot 10^{-18}$\,m$^{2}$. The only free parameter, $T_0$, is chosen by optimizing the agreement of the mode frequencies to the experimental results. We obtain $T_0 = 0.1$\,nN. The resulting $z_s$ as a function of $V_g$ is shown in Fig.~2(e) of the main text. Finite element calculations with a straight tube yield values of $z_s$ close to the results of Eq.~\ref{EulerBernoulli_5}, see dashed line in Fig.~2(e) of the main text. The resonance frequencies of the six first modes are depicted in Fig.~\ref{SM12}(a).

\section{Euler-Bernoulli equation: coupling between modes}
In the previous section, we have disregarded terms that couple different modes. The coupling has the same origin as the quadratic and cubic nonlinearities ($\alpha_2$ and $\alpha_3$). It arises from the tension that is induced in a mode when another mode oscillates (through the integral term $\int_0^L \left(\frac{d z}{d x} \right)^2 dx$ in Eq.~\ref{EulerBernoulli_1}). As a consequence, the coupling coefficients are of the same order of magnitude as $\alpha_2$ and $\alpha_3$. Assuming that $z(x,t) = z_s \phi_s(x) + z_1 \phi_1(x) + z_2 \phi_2(x)$, Eq.~\ref{EulerBernoulli_1} leads to

\begin{equation}
\frac{d^2z_1(t)}{dt^2} = - \omega_0^2 z_1(t) - \alpha_2 z_1^2(t) - \alpha_3 z_1^3(t) - \beta_{22} z_2^2(t) - \epsilon_{122} z_1(t) z_2^2(t) \label{EulerBernoulli_14}
\end{equation}

where $\omega_0$, $\alpha_2$, and $\alpha_3$ are given by the expressions in Eq.~\ref{EulerBernoulli_4}, $\beta_{22} = \frac{E}{4 \rho} \left(\frac{\pi}{L} \right)^4 z_s$, and $\epsilon_{122} = \frac{E}{4 \rho} \left(\frac{\pi}{L} \right)^4$ (all other summands $\beta_{jk}$ and $\epsilon_{jkl}$ that appear in Eq.~1 of the main text are zero). The equation of motion for the lowest mode moving parallel to the gate electrode is

\begin{equation}
\frac{d^2z_2(t)}{dt^2} = - \omega_0^2 z_2(t) - \alpha_3 z_2^3(t) - \beta_{12} z_1(t) z_2(t) - \epsilon_{112} z_2(t) z_1^2(t) \label{EulerBernoulli_15}
\end{equation}

where $\omega_0$ and $\alpha_3$ are given by Eq.~\ref{EulerBernoulli_7} and the coupling coefficients are $\beta_{12} = \frac{E}{2 \rho} \left(\frac{\pi}{L} \right)^4 z_s$ and $\epsilon_{112} = \frac{E}{4 \rho} \left(\frac{\pi}{L} \right)^4$.

\section{Onset of nonlinearity}
Nonlinear effects set in when $z_0$ reaches a critical value $z_c$. From Ref.~\cite{Lifshitz2008S}, we infer $z_c = 1.24 \omega_0 / \sqrt{Q \left|\alpha \right|}$ assuming nonlinear damping to be negligible. Here, $\alpha$ is the effective nonlinear coefficient (see next section) that can be extracted from the backbone function connecting all resonance peaks at different driving amplitudes. From Ref.~\cite{Lifshitz2008S}, we have

\begin{equation}
\omega_{max} - \omega_0 = \frac{3}{8} \frac{\alpha z_0^2}{\omega_0}, \label{nonlin_1}
\end{equation}

where $\omega_{max} / 2 \pi$ is the frequency where the amplitude is largest and $\omega_0 / 2 \pi$ is the resonance frequency in the linear regime. In Fig.~1(d) of the main text, we have $\omega_{max} = 2 \pi \cdot 46.85$\,MHz, $\omega_0 = 2 \pi \cdot 46.35$\,MHz, $z_0 = 8.7$\,nm, and thus get $\alpha = 3.2 \cdot 10^{31}$\,m$^{-2}$s$^{-2}$. Using $Q = 230$ extracted from the resonance width in Fig.~1(c) of the main text, this leads to a critical amplitude of $z_c = 4.2$\,nm, which is consistent with $z_0 = 3.2$\,nm in Fig.~1(c) (where the resonance displays no hysteresis) and with $z_0 = 8.7$\,nm in Fig.~1(d) (where there is hysteresis). We repeat the  same procedure for the data in Fig.~1(f) of the main text. Here, $\omega_{max} = 2 \pi \cdot 123.5$\,MHz, $\omega_0 = 2 \pi \cdot 124.25$\,MHz, $z_0 = 2.1$\,nm, and we get $\alpha = -2.2 \cdot 10^{33}$\,m$^{-2}$s$^{-2}$. With $Q = 354$, we calculate $z_c = 1.1$\,nm, which again is consistent with the results in Fig.~1(e) and (f) ( where $z_0 = 0.9$\,nm and $2.1$\,nm, respectively).

\section{Static displacement}
The reversal of the asymmetry of the resonance between Fig.~1(d) and (f) of the main text is due to a sign change of the effective nonlinearity $\alpha$~\cite{Kozinsky2006S,Lifshitz2008S} which depends on the quadratic and cubic coefficients in Eq.~\ref{EulerBernoulli_6} as~\cite{Nayfeh1979S}

\begin{equation}
\alpha = \alpha_3 - \frac{10}{9} \omega_0^{-2} \alpha_2^2. \label{nonlin_2}
\end{equation}

A comparison of Eq.~\ref{EulerBernoulli_4} and Eq.~\ref{EulerBernoulli_6} reveals that

\begin{equation}
\alpha_2 = \frac{3 E}{4 \rho} z_s \left(\frac{\pi}{L} \right)^4 \label{nonlin_3}
\end{equation}

and

\begin{equation}
\alpha_3 = \frac{E}{4 \rho} \left(\frac{\pi}{L} \right)^4. \label{nonlin_4}
\end{equation}

The asymmetry of the resonance can be used to estimate the static displacement $z_s$ of the resonator at different values of $V_g$. With $\alpha$ known from the estimations in the last section, we can insert Eq.~\ref{nonlin_3} and Eq.~\ref{nonlin_4} into Eq.~\ref{nonlin_2} in order to obtain $z_s$. With $E = 1$\,TPa, $\rho = 2300$\,kgm$^{-3}$, and $L = 1.77\,\mu$m, we get $z_s = 2.8$\,nm for $V_g = 1.5$\,V, and $z_s = 13$\,nm for $V_g = 4$\,V.

\section{Electrostatic nonlinearities}
In the previous sections, we have shown that the sign change of the effective nonlinearity $\alpha$ (from a positive to a negative value) as a function of $V_g$ is consistent with the expected increase of the quadratic nonlinearity $\alpha_2$. A negative $\alpha$ could also have an electrostatic origin~\cite{Kozinsky2006S,Lifshitz2008S}. The quadratic and cubic nonlinearities are $\alpha_2^{el} = - \frac{1}{2 m} C''' V_g^2$ and $\alpha_3^{el} = - \frac{1}{2 m} C'''' V_g^2$, where $m$ is the effective mass of the resonator and $C'''$ and $C''''$ are the third and fourth derivatives of the capacitance with respect to displacement. The values calculated for $V_g = 4$\,V are $\alpha_2^{el} = - 4.4 \cdot 10^{22}$\,m$^{-1}$s$^{-2}$ and $\alpha_3^{el} = - 4.0 \cdot 10^{29}$\,m$^{-2}$s$^{-2}$, which are at least $3$ orders of magnitude smaller than the mechanical nonlinearities above. We therefore neglect electrostatic nonlinearities in the analysis of our experiment.

\begin{figure*}
\includegraphics[width=172mm]{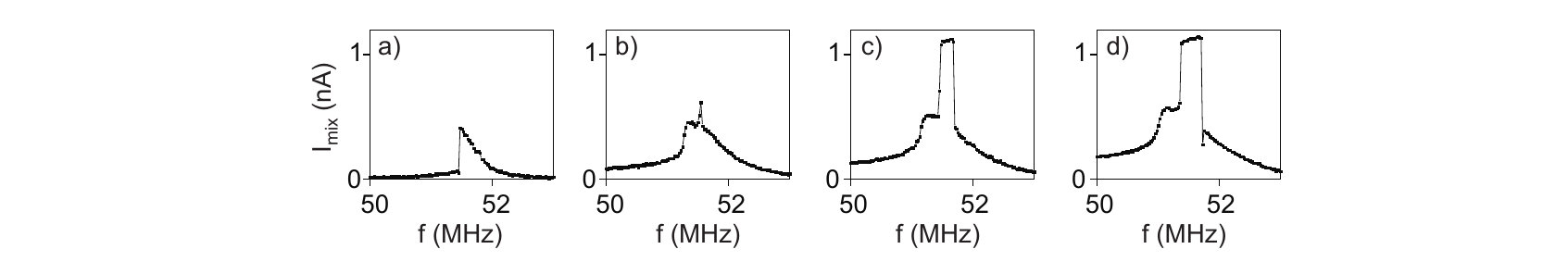}
\caption{\label{SM11} Resonance lineshape for different driving forces. The resonance is measured with the two-source method for $V_g = 1.88$\,V and $V_{s}^{ac} = 0.28$\,mV. $V^{ac} = 0.56$\,mV in (a), $V^{ac} = 2.2$\,mV in (b), $V^{ac} = 3.4$\,mV in (c), and $V^{ac} = 4.5$\,mV in (d). A conventional Duffing nonlinearity (with $\alpha < 0$) is recovered at the lowest driving force. Sweeps are performed with decreasing frequency.}
\end{figure*}

\begin{figure*}
\includegraphics[width=172mm]{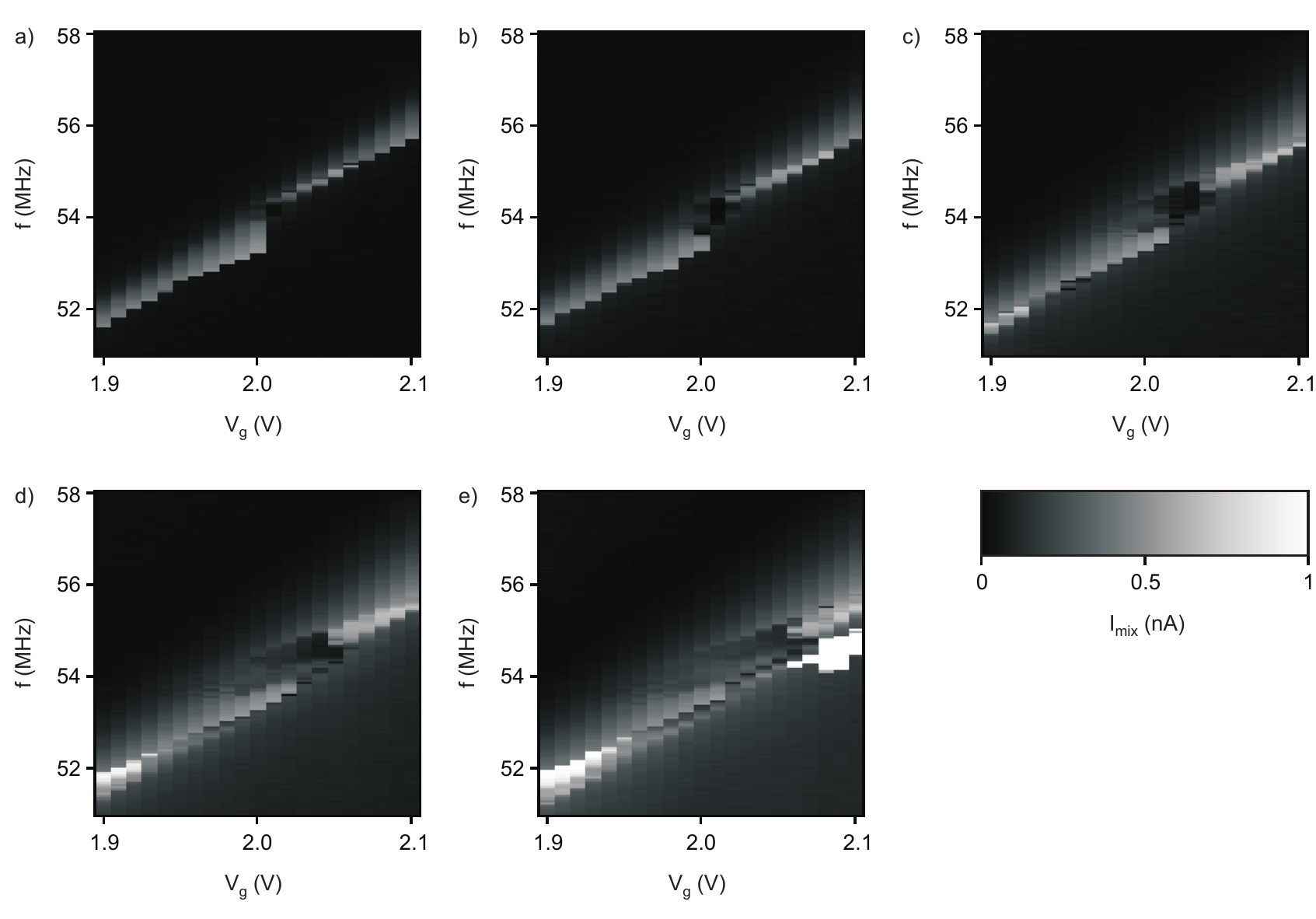}
\caption{\label{SM7} Mixing current as a function of $f$ and $V_g$ for different driving forces. $I_{mix}$ is measured with the two-source technique with $V_{s}^{ac} = 0.28$\,mV. $V^{ac} = 0.56$\,mV in (a), $V^{ac} = 1.1$\,mV in (b), $V^{ac} = 2.2$\,mV in (c), $V^{ac} = 3.4$\,mV in (d), and $V^{ac} = 4.5$\,mV in (e). Sweeps are performed with decreasing frequency. Panel (b) corresponds to Fig.~4(f) of the main text.}
\end{figure*}

\begin{figure*}
\includegraphics[width=172mm]{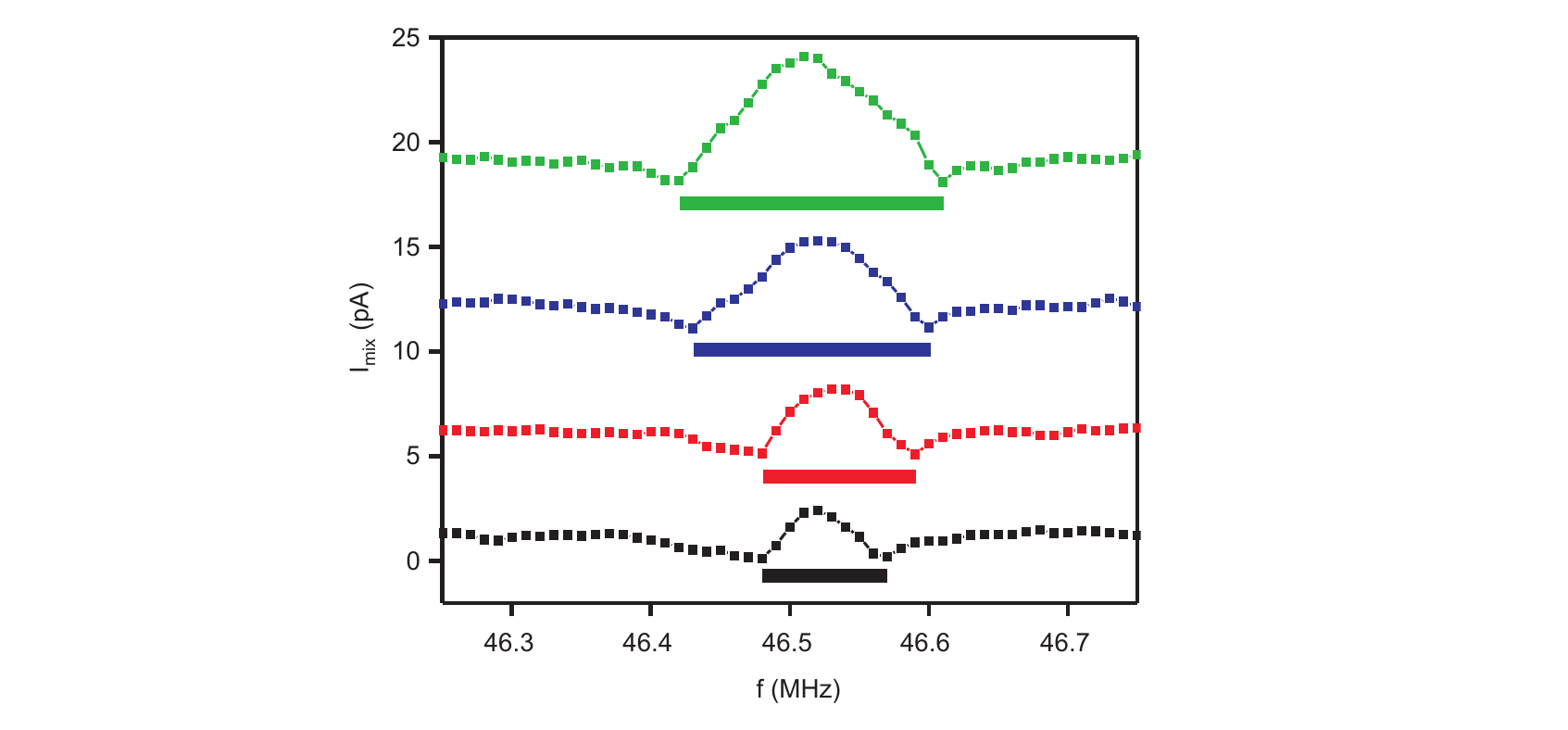}
\caption{\label{SM9} Increase of the resonance width with $V^{ac}$. We measure the resonance of mode $N$ at $V_g = 1.5$\,V with the FM technique for $V^{ac} = 0.1$\,mV (black), $V^{ac} = 0.12$\,mV (red), $V^{ac} = 0.15$\,mV (blue), and $V^{ac} = 0.2$\,mV (green). The distance between the two minima flanking the resonance peak (solid bars) corresponds to the resonance width $\Delta f = f_0 / Q$.}
\end{figure*}

\begin{figure*}
\includegraphics[width=172mm]{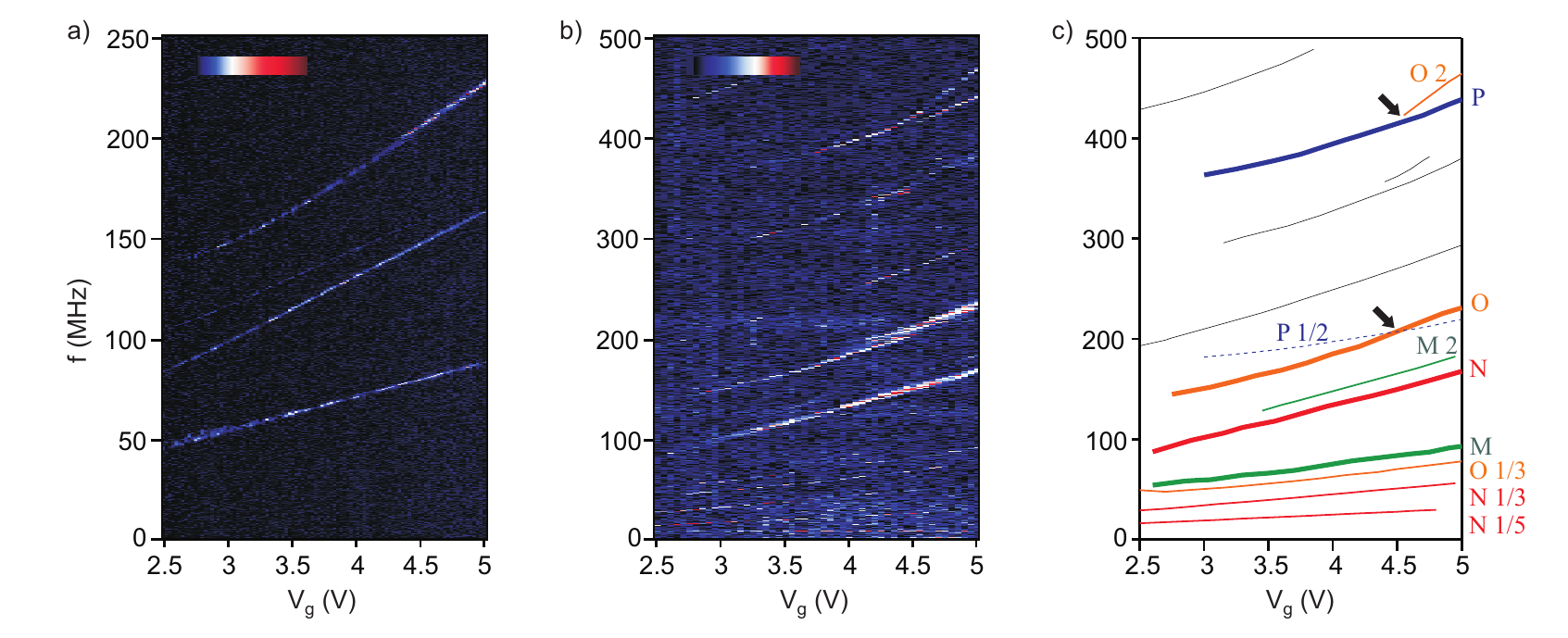}
\caption{\label{SM4} Map of resonance frequencies as a function of $V_g$. (a) Frequency modulation measurement at low driving force (obtained by measuring $I_{mix}$ as a function of $f$ and $V_g$ with $V^{ac}=4$\,mV). Three resonances are detected with a large signal, a fourth one shows up faintly. Colour scale: $0$ (black) to $0.1$\,nA (dark red). (b) Same measurement with a larger frequency range and $V^{ac}=40$\,mV. Colour scale: $0$ (black) to $0.1$\,nA (dark red). (c) Schematic of the map of the resonance frequencies as a function of $V_g$. Resonances that cannot be assigned to a mode are drawn in black. Black arrows mark regions where modes $O$ and $P$ are commensurate and the resonances lineshapes become exotic.}
\end{figure*}

\begin{figure*}
\includegraphics[width=172mm]{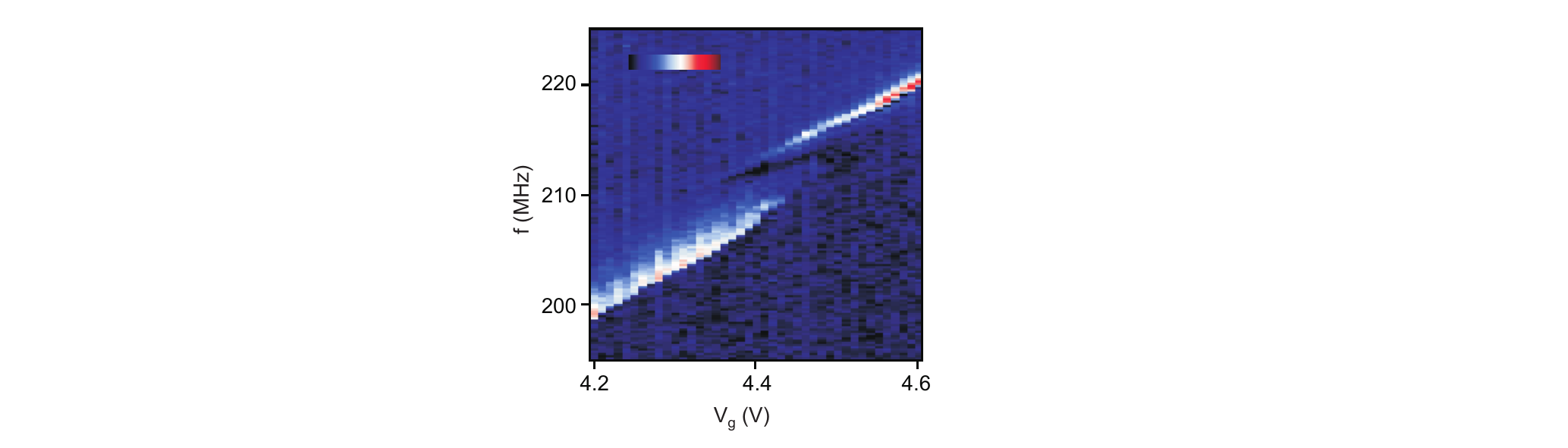}
\caption{\label{SM15} Modal coupling in the second device. Measurement with the two-source method (obtained by measuring $I_{mix}$ as a function of $f$ and $V_g$ with $V^{ac}=2.8$\,mV and $V_s^{ac}=0.28$\,mV). Colour scale: $0$ (black) to $0.1$\,nA (dark red).}
\end{figure*}

\section{Resonance lineshapes}
In Fig.~4 of the main manuscript, we show resonance lineshapes with exotic features that we associate with mechanical coupling between commensurate modes of the nanotube. Here, we show how these features vanish and how a usual Duffing nonlinearity is recovered when we reduce the driving force. The different panels in Fig.~\ref{SM11} correspond to the same resonance for different driving forces ($\propto V^{ac}$).

A second example with increasing driving force is displayed in Fig.~\ref{SM7}. Here, we can observe the evolution from an almost unbroken line (resonance frequency versus $V_g$) in Fig.~\ref{SM7}(a) to a highly exotic response with multiple peaks and dips as a function of $f$ in Fig.~\ref{SM7}(e).

\section{Nonlinear damping}
We recently reported that the resonance width $\Delta f = f_0 / Q$ of nanotube and graphene mechanical resonators can depend on the driving voltage $V^{ac}$. We attributed this phenomenon to the nonlinear damping force $\eta z^2 \dot{z}$~\cite{Eichler2011S}. In the present device, the dominant bistability behaviour prevents observing nonlinear damping above $V^{ac} \sim 0.2$\,mV. Nontheless, we found an increase of the resonance width for $V^{ac} \leq 0.2$\,mV (Fig.~\ref{SM9}). In this measurement, we use the frequency modulation technique, which produces two characteristic minima flanking the resonance peak. The separation of these minima corresponds to $\Delta f$. A clear increase of $\Delta f$ is seen between $V^{ac} = 0.1$\,mV (below this driving voltage the signal vanishes) and $V^{ac} = 0.2$\,mV.

\section{Additional device}
A second nanotube resonator exhibits similar behaviour as the one discussed so far. The results of the second device are summarized in Fig.~\ref{SM4} and Fig.~\ref{SM15}. Three modes are clearly visible at low driving force, while a fourth resonance shows up faintly [Fig.~\ref{SM4}(a)]. In these measurements, we use the FM technique with $V^{ac} = 4$\,mV. With a larger driving force ($V^{ac} = 40$\,mV), many more resonances appear [Fig.~\ref{SM4}(b)]. With this device, we have mapped the frequency spectrum up to $500$\,MHz, and all detected resonances are depicted in Fig.~\ref{SM4}(c) and labelled according to the most probable harmonic spectrum. We can identify regions where two modes are commensurate or nearly commensurate and the resonance lineshapes become exotic (black arrows). Around $V_g = 4.4$\,V, mode $P$ has exactly twice the frequency of mode $O$ (Fig.~\ref{SM15}). There, we observe a discontinuity in the map of the resonance frequency as a function of $V_g$.

\end{document}